\documentclass[a4paper,fleqn]{cas-dc}

\usepackage[numbers]{natbib}
\usepackage{graphicx,amssymb,amsmath,booktabs,multirow,xcolor,subcaption,bm}
\newcommand{\iqr}[3]{$#1\ [#2\text{--}#3]$}

\def\tsc#1{\csdef{#1}{\textsc{\lowercase{#1}}\xspace}}
\tsc{WGM}
\tsc{QE}
\tsc{EP}
\tsc{PMS}
\tsc{BEC}
\tsc{DE}

\begin{document}
\let\WriteBookmarks\relax
\def\floatpagepagefraction{1}
\def\textpagefraction{.001}
\shorttitle{CT-to-MRI pretraining transfer for rectal tumor segmentation}
\shortauthors{A. Rangnekar et~al.}

\title [mode = title]{Tumor-aware augmentation with task-guided attention analysis improves rectal cancer segmentation from magnetic resonance images}

\tnotemark[1]
\tnotetext[1]{This study was supported by the Simons Foundation and the Breast Cancer Research Foundation (through grant MATH-23-001), and the NIH ROBIN cooperative group (grant U54CA274291). This work utilized resources from the High-Performance Computing Group at Memorial Sloan Kettering Cancer Center.}
\author[1]{Aneesh Rangnekar}[type=editor,
                        orcid=0000-0002-0079-9495]
\credit{Conceptualization, Methodology, Software, Validation, Formal analysis, Investigation, Data curation, Writing - Original Draft, Writing - Review \& Editing, Visualization}

\author[2]{Joao Miranda}
\credit{Data Curation, Writing - Review \& Editing}

\author[2]{Natally Horvat}
\credit{Data Curation, Writing - Review \& Editing, Supervision}

\author[2]{Stephanie Chahwan}
\credit{Data Curation}

\author[2]{Samir Alrayess}
\credit{Data Curation}

\author[1]{Aditya Apte}
\credit{Investigation, Software, Resources}

\author[1]{Aditi Iyer}
\credit{Investigation, Software, Resources}

\author[1]{Eve LoCastro}
\credit{Investigation, Resources}

\author[3]{Revathi Ravella}
\credit{Validation, Data Curation}

\author[2]{Marc J Gollub}
\credit{Validation, Writing - Review \& Editing}

\author[2]{Iva Petkovska}
\credit{Validation, Writing - Review \& Editing}

\author[4]{Jesse Joshua Smith}
\credit{Validation, Writing - Review \& Editing, Supervision}

\author[3,5]{Paul Romesser}
\credit{Validation, Writing - Review \& Editing, Supervision}

\author[4]{Julio Garcia-Aguilar}
\credit{Validation, Writing - Review \& Editing, Supervision}

\author[1]{Harini Veeraraghavan}
\credit{Conceptualization, Methodology, Data Curation, Writing - Original Draft, Writing - Review \& Editing, Supervision, Project administration}
\cormark[1]
\ead{veerarah@mskcc.org}

\author[1]{Joseph O Deasy}
\credit{Conceptualization, Methodology, Investigation, Writing - Review \& Editing, Supervision, Project administration, Funding acquisition}

\cortext[cor1]{Corresponding author}

\affiliation[1]{organization={Department of Medical Physics, Memorial Sloan Kettering Cancer Center},
                addressline={1275 York Avenue},
                city={New York},
                state={NY},
                postcode={10065},
                country={USA}}

\affiliation[2]{organization={Department of Radiology, Memorial Sloan Kettering Cancer Center},
                addressline={1275 York Avenue},
                city={New York},
                state={NY},
                postcode={10065},
                country={USA}}

\affiliation[3]{organization={Department of Radiation Oncology, Memorial Sloan Kettering Cancer Center},
                addressline={1275 York Avenue},
                city={New York},
                state={NY},
                postcode={10065},
                country={USA}}

\affiliation[4]{organization={Department of Surgery, Memorial Sloan Kettering Cancer Center},
                addressline={1275 York Avenue},
                city={New York},
                state={NY},
                postcode={10065},
                country={USA}}
                
\affiliation[5]{organization={Department of Medicine, Memorial Sloan Kettering Cancer Center},
                addressline={1275 York Avenue},
                city={New York},
                state={NY},
                postcode={10065},
                country={USA}}
                
\begin{abstract}
Although self-supervised pretraining is expected to learn broadly transferable representations, its effectiveness across imaging modalities substantially different from the pretraining domain, and on complex tumor-segmentation tasks, remains understudied. Evaluating CT-pretrained transformers on MRI rectal cancer segmentation, we identified two interacting failure modes in CT-to-MRI transfer: (a) inefficient token usage caused by zero-padding to match pretrained input dimensions, and (b) ineffective feature adaptation. We investigated these vulnerabilities using two primary CT-pretrained hierarchical shifted-window transformer backbones, SMIT and Swin~UNETR, together with VoCo as a large-scale-pretrained supporting benchmark; these models differ in pretraining objectives and datasets. Mechanistic analysis leveraged an attention dilution index (ADI), an entropy-based metric quantifying attention diverted toward uninformative padding tokens, and centered kernel alignment (CKA) to measure feature reuse during MRI adaptation. ADI increased with zero-padding, while high feature reuse did not necessarily translate to improved downstream accuracy. To mitigate these issues, we introduced two interventions: a tumor-aware augmentation strategy to expand tumor appearance heterogeneity coverage, and an anisotropic cropping strategy to restore token efficiency. Fine-tuning with these strategies on identical rectal MRI datasets yielded detection rates of 91.1\% (225/247) and 88.7\% (219/247) for the primary SMIT and Swin~UNETR backbones, with the supporting VoCo benchmark reaching 90.3\% (223/247), demonstrating significantly improved robustness under CT-to-MRI transfer. This study is among the first to examine when pretrained transformers fail to transfer across imaging modalities and demonstrates how targeted mitigation strategies, motivated by mechanistic analysis, can systematically overcome cross-modality transfer limitations.
\end{abstract}



\begin{keywords}
Rectal cancer, tumor segmentation, Swin Transformer, attention dilution, data augmentation, transfer learning, medical imaging.
\end{keywords}

\maketitle

\section{Introduction}
\label{sec:intro}

Colorectal cancer is the third most common malignancy worldwide, with rectal cancer accounting for approximately one-third of cases and an estimated 45,000 new diagnoses annually in the United States alone~\cite{siegel2022cancer,siegel2026leading}. T2-weighted MRI, often interpreted in conjunction with diffusion-weighted imaging, is commonly used for diagnosis, staging, and longitudinal response assessment in rectal cancer because of its superior soft-tissue contrast compared to computed tomography scans~\cite{battersby2016prospective,beets2018magnetic,vliegen2008accuracy}. Accurate volumetric delineation of rectal tumors on MRI is required for volume-based radiomics and response assessment. However, manual contouring is laborious and subject to inter-observer variation, impacting reliability of scan-derived radiomics features~\cite{Charbel2025RectalMRIRadiomics,Tixier2019ReliabilitySegmentationGBM}. Prior work on automated rectal MRI tumor segmentation remains insufficiently accurate~\cite{lin2023fully,yang202421}, motivating our approach.

Transformer-based segmentation models that combine self-attention with convolutional decoding~\cite{chen2024transunet,dosovitskiy2021,hatamizadeh2021swin,jiang2022self,tang2022self} have recently emerged as strong alternatives to convolutional-only architectures~\cite{hamabe2022artificial,isensee2021nnu,knuth2022semi,milletari2016v,ronneberger2015u,trebeschi2017deep}, due to their ability to model long-range spatial context using multi-head self-attention, as well as data-efficient fine-tuning enabled by pretraining of transformer backbones~\cite{matsoukas2021time,raghu2019transfusion,tang2022self}. Because CT datasets are more abundant and span multiple disease sites, pretraining is often performed using these datasets and then applied to MRI under the assumption that spatial representations learned from large CT corpora will transfer seamlessly across modalities~\cite{chen2019med3d,jiang2022self,tang2022self}.

Whether this assumption holds in practice, however, remains poorly understood. In rectal T2-weighted (T2W) MRI segmentation, we find that cross-modal transfer from CT-pretrained transformers may be limited by two distinct and interacting failure modes. First, rectal cancers exhibit substantial tumor appearance heterogeneity on MRI that is challenging to capture by models pretrained using CT datasets alone. The second is token inefficiency arising from geometric mismatch: pelvic MRI volumes are typically anisotropic along the \textit{z} dimension and contain substantially fewer slices than the isotropic input sizes expected by pretrained transformer backbones, necessitating extensive zero-padding. Unlike learned mechanisms such as attention sinks or registers, which can stabilize attention routing~\cite{darcet2023vision,xiao2023streamingllm}, the padding tokens in this setting contain no useful signal and instead may consume attention-routing capacity that dilute informative attention allocation.

To understand the source of these failure modes, we performed a mechanistic analysis focusing on token usage as well as feature reuse. We introduced an attention dilution index (ADI), an entropy-based metric to quantify the extent to which attention routing is diverted towards uninformative padding tokens. Feature reuse was analyzed using centered kernel alignment (CKA)~\cite{cortes2012algorithms,kornblith2019similarity}. Our analysis shows that zero-padding progressively consumes routing capacity in hierarchical vision transformers and that reducing the padding burden restores useful computation. To mitigate these failure modes, we introduced a tumor-aware data augmentation strategy that selectively perturbs intensity statistics within tumors to increase the range of tumor appearances observed during training. We also found that decreasing the input depth during fine-tuning while preserving the pretrained backbone weights reduced zero-padding inefficiency and lowered computational cost by 56\%. Together, these findings clarify why CT-pretrained transformers may transfer poorly to rectal MRI and provide practical strategies for improving accuracy of models fine-tuned to out-of-distribution (OOD) modalities.

\section{Methods}
\label{sec:methods}

\subsection{Study cohort}
\label{sec:methods:study_cohort}

The training cohort was sourced from two retrospective datasets from our institution previously used for radiomics and radiogenomic analysis~\cite{horvat2018mr,petkovska2020clinical,miranda2023mri} with institutional guidelines, applicable regulations, and the Declaration of Helsinki. The study was approved by the Institutional Review Board of Memorial Sloan Kettering Cancer Center. A waiver of informed consent was granted due to the retrospective nature of the study and the use of de-identified imaging data. All patient data were handled in accordance with institutional privacy policies, and the privacy rights of human subjects were protected throughout the study.

The training dataset consisted of 169 pre-treatment high-resolution oblique axial T2-weighted MRI volumes of patients with biopsy proven locally advanced rectal adenocarcinoma collected between 2009 and 2016. Three abdominal radiologists, with $\geq$ 3 years of experience in rectal MRI, reviewed all images and reached a consensus on the tumor location~\cite{nougaret2013use}. One of the radiologists then manually segmented the tumor in all slices on the high spatial resolution axial oblique T2WI using a free open-source software package (ITK-SNAP, version 3.4.0), to provide the volume of interest of the tumor for radiomics-based quantitative image analysis. Cases with incomplete tumor coverage on MRI, poor image quality and mucinous tumors were excluded from analysis. MRIs were performed either on 1.5 Tesla or 3 Tesla GE Healthcare systems using phased-array coils, with axial oblique T2-weighted sequences acquired perpendicular to the tumor axis (typical slice thickness 2 to 4 mm, field of view $\sim$180 to 220~mm). An independent retrospectively collected held-out test cohort of $N=247$ patients from the prospective Organ Preservation in Rectal Adenocarcinoma (OPRA) clinical trial was accessed after the model was trained and locked for testing~\cite{garcia2022organ,miranda2023mri}. Pre-treatment T2-weighted MRI volumes acquired on a Signa HDx 1.5 Tesla GE scanner with high-resolution thin slices (3 mm) were used for analysis. 

All design choices pertaining to experiment protocols (anisotropic crop size, data augmentation, fine-tuning hyperparameters), and preprocessing (intensity clipping/normalization, crop geometry), were determined using the training/development cohort through 3-fold cross-validation. Cohort characteristics are summarized in Table~\ref{tab:cohort_summary}.

For our experiments, all MRI scans were resampled to isotropic 1 mm$^3$ resolution to ensure consistent tokenization with the pretrained backbone. Image intensities were clipped at the 99$^\text{th}$ percentile and normalized to $[0,1]$, reducing scanner-dependent outliers while preserving relative contrast. Image volumes were resampled using bilinear interpolation, while segmentation masks were resampled using nearest-neighbor interpolation. 

\begin{table}[t]
\def\arraystretch{1.2}
\centering
\caption{Cohort characteristics. In-plane resolution and native axial extent after resampling are reported as median [IQR]. Padding-fraction statistics under the cubic input crops are reported separately in Section~\ref{sec:results_padding_adi}.}
\label{tab:cohort_summary}
\resizebox{0.98\columnwidth}{!}{%
\begin{tabular}{lll}
 & Training & Test \\
 \midrule
$N$ (scans) & 169 & 247 \\
In-plane resolution (mm) & \iqr{0.43}{0.39}{0.47} & \iqr{0.43}{0.35}{0.63} \\
Slice thickness (mm) & 3.00--4.00 & 3.00--4.00 \\
\begin{tabular}[c]{@{}l@{}}Native axial extent\\ after resampling\end{tabular} & \iqr{88}{70}{109} & \iqr{93}{70}{120} \\
\bottomrule
\end{tabular}%
}
\end{table}

\subsection{Model architecture}
\label{sec:methods:architecture}

We used a hybrid transformer–convolution architecture consisting of a pretrained Swin Transformer encoder and a convolutional U-Net decoder~\cite{ronneberger2015u} connected through multi-scale skip connections. Two Swin-based backbones were evaluated:

\begin{itemize}
    \item \textbf{SMIT}, pretrained in-house via masked image prediction and token self-distillation on a large CT corpus of 10,432 scans, using 128 $\times$ 128 $\times$ 128 input size with patch size 2 $\times$ 2 $\times$ 2 and window size 4 $\times$ 4 $\times$ 4.
    \item \textbf{Swin~UNETR}, initialized from publicly released NVIDIA pretrained weights derived from 5,050 CT scans using image inpainting and contrastive objectives, with 96 $\times$ 96 $\times$ 96 input size, patch size 2 $\times$ 2 $\times$ 2, and window size 7 $\times$ 7 $\times$ 7.
\end{itemize}

These two backbones were selected because they provide a controlled comparison within the same architecture family; both are 3D hierarchical shifted-window transformer encoders coupled to U-Net-style decoders, both require fixed 3D input crops, and both have CT-pretrained weights available for transfer to downstream segmentation. At the same time, they differ in pretraining objective, weight source, input size, window size, and depth, allowing us to test whether the observed failure modes were specific to a single implementation or present across related Swin-based 3D transformer backbones~\cite{hatamizadehSwinUNETR2022,jiang2022self,liu2021swin,tang2022self}. We therefore interpret our findings as strongest for fixed-crop hierarchical windowed transformers under CT-to-MRI transfer, rather than as a universal claim about all transformer architectures.

\subsection{Tumor-aware intensity augmentation}
\label{sec:methods:tumor_aug}

\begin{figure}
\centering
\includegraphics[width=0.98\linewidth]{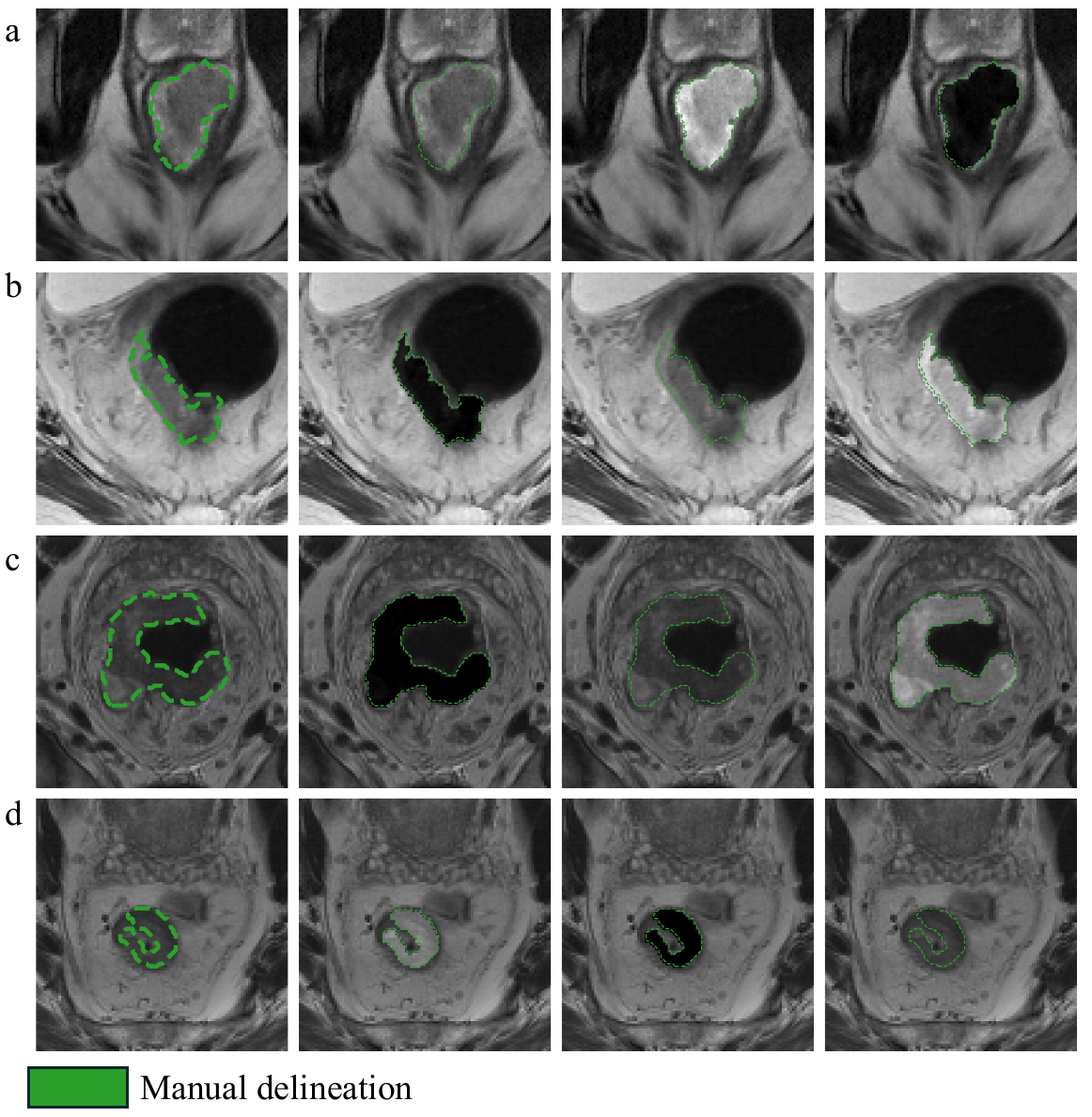}
\caption{Tumor-aware intensity augmentation applied to representative training cases. Each row shows an axial slice with manual delineation (green), followed by three augmented variants in which tumor-region intensities are randomly scaled and shifted.}
\label{fig:tumor_aware_augmentation}
\end{figure}

Standard intensity augmentation transforms the full image uniformly and therefore may inadequately capture localized inter-subtype appearance differences, discussed further in Section~\ref{sec:methods:clustering_subtypes}. We implemented a tumor-localized intensity perturbation, where $M(x)$ denotes the tumor mask:

\begin{equation}
I'(x) = 
\begin{cases}
\alpha \cdot I(x) + \beta & \text{if } M(x) > 0 \\ I(x) & \text{otherwise.}
\end{cases}
\end{equation}

Here $\alpha \sim \mathcal{U}(0.5,1.5)$ and $\beta \sim \mathcal{U}(-0.2,0.2)$. These ranges were selected from training/development cohort tumor-intensity statistics, including the observed bright and dark subtype contrast ranges, to broaden local tumor appearance while remaining bounded enough to avoid implausible contrast inversions or alteration of background anatomy. The perturbation was applied with probability $0.3$ after spatial augmentations during training only, so that tumor-localized appearance variation was increased without replacing the original intensity distribution. Figure~\ref{fig:tumor_aware_augmentation} illustrates that the transform alters tumor appearance locally while leaving surrounding anatomy unchanged.

\subsection{Attention dilution analysis}
\label{sec:methods:token_analysis}

Rectal MRI volumes are typically acquired with relatively thick slices and limited $z$-plane extent, producing anisotropic volumes that fall short of the cubic input sizes expected by pretrained transformer backbones. Matching these input requirements therefore necessitates extensive zero-padding, introducing large numbers of uninformative tokens into every self-attention computation.

In standard scaled dot-product self-attention, query and key vectors first interact to produce attention weights as

\begin{equation}
\alpha_{ij} = 
\frac{\exp(q_i^\top k_j / \sqrt{d})}
{\sum_{m} \exp(q_i^\top k_{m} / \sqrt{d})},
\qquad
\sum_j \alpha_{ij} = 1.
\label{eq:attention_weights}
\end{equation}
where $q_i$ and $k_j$ are the query and key vectors for tokens $i$ and $j$ respectively, and $d$ is the key dimension. These weights are then used to aggregate value vectors into the output:

\begin{equation}
o_i = \sum_{j} \alpha_{ij} v_j.
\label{eq:attention_output}
\end{equation}

Critically, the softmax normalization aggregates contributions from \textit{all} tokens within the attention window, including both real tokens and padding tokens. Since the total attention budget for any given token is fixed at 1 (Equation~\ref{eq:attention_weights}), any focus assigned to padding tokens represents a direct loss of informative attention capacity. In essence, the padding acts as a distraction that can dilute the model's ability to represent actual anatomical structures. To mitigate this effect, we adopt an anisotropic crop of 128 $\times$ 128 $\times$ 64 pixels during fine-tuning while preserving the pretrained backbone weights. The 64-slice depth was selected as a geometry- and architecture-driven compromise rather than as a test-set-optimized continuous hyperparameter: among discrete crop depths compatible with hierarchical patch merging, windowed attention, and sliding-window inference, it substantially reduces through-plane padding in rectal MRI while preserving the in-plane field of view and tumor-centered anatomical context. A quantitative padding-versus-depth analysis supporting the 64-slice choice is provided in Table~\ref{tab:crop_depth}, and the computational footprint of both interventions is summarized in Table~\ref{tab:compute}. This modification substantially reduces padding tokens and lowers computational cost by 56\%. To quantify this effect, we introduce the \textit{attention dilution index}, a diagnostic metric designed to measure the degree of attention dilution post hoc, rather than to drive architectural decisions.

For our analysis, we define the padding fraction, denoted as $pf$, as
\begin{equation}
pf = \max\!\left(1 - \frac{L}{C},\ 0\right),
\label{eq:padding_fraction}
\end{equation}
where $L$ is the native $z$-extent of the volume after resampling and $C$ is the crop size along $z$ (128 for SMIT; 96 for Swin~UNETR under cubic crops, 64 for both under ACT). When $L \geq C$, the volume is cropped to $C$ and $pf = 0$; when $L < C$, the shortfall is filled with zero-padding and $pf > 0$.

Formally, let $\mathcal{R}_S$ and $\mathcal{P}_S$ denote the sets of real and padding tokens at transformer stage $S$. The softmax normalization for each query token can therefore be decomposed into interactions between real and padding tokens:

\begin{equation}
\sum_{j \in \mathcal{R}_S} \alpha_{ij} +
\sum_{j \in \mathcal{P}_S} \alpha_{ij} = 1 .
\end{equation}

Considering both real and padding queries, this partition yields four interaction types: $R \rightarrow R$, $R \rightarrow P$, $P \rightarrow R$, and $P \rightarrow P$. In this work, we focus on the real query tokens ($i \in \mathcal{R}_S$), since padding queries do not correspond to anatomical signal and are therefore excluded from analysis. Token real/padding status was propagated through patch-merging stages using max-pooling over the binary mask: a merged token was classified as real if any of its constituent tokens were real. 

Specifically, for each real query token $i$ with post-softmax attention weights $\{\alpha_{ij}\}$, we decompose the attention entropy into contributions from real and padding keys:

\begin{equation}
\begin{aligned}
H_i^{R \rightarrow R} = -\sum_{j \in \mathcal{R}_S} \alpha_{ij}\log\alpha_{ij}, \\
H_i^{R \rightarrow P} = -\sum_{j \in \mathcal{P}_S} \alpha_{ij}\log\alpha_{ij}.
\end{aligned}
\end{equation}

\noindent The attention dilution index at stage $S$ is defined as:

\begin{equation}
\mathrm{ADI}_S =
\frac{\displaystyle\sum_{i \in \mathcal{R}_S} H_i^{R \to P}}
{\displaystyle\sum_{i \in \mathcal{R}_S} H_i^{R \to R} +
\displaystyle\sum_{i \in \mathcal{R}_S} H_i^{R \to P}},
\label{eq:adi}
\end{equation}
where $\mathrm{ADI}_S$ ranges from 0 (no routing uncertainty lost to padding) to 1 (all routing uncertainty consumed by padding tokens). Statistics were averaged across attention heads, windows, transformer blocks, and scans. Intuitively, $\mathrm{ADI}_S$ measures the fraction of attention routing uncertainty that is expended on padding tokens rather than real tokens. We intentionally defined ADI in terms of entropy to measure how much routing uncertainty is expended on zero-padded regions during token aggregation. This definition allows us to test our hypothesized failure mode that consumption of routing capacity by padding tokens blocks otherwise useful information from being allocated among informative tokens.

\subsection{Training and evaluation protocol}

\subsubsection{Fine-tuning hyperparameters}
\label{sec:methods:finetuning_details}

All models were fine-tuned using an equally weighted combination of soft Dice and cross-entropy loss, optimized with AdamW (learning rate $3 \times 10^{-4}$, weight decay $10^{-5}$) with a linear warmup cosine annealing schedule (50 epoch warmup). Pretrained models were trained for 500 epochs and randomly initialized models for 1,000 epochs, with batch size 2 distributed across 4 NVIDIA A100 GPUs. The best checkpoint was selected based on the highest mean Dice on the validation split. Inference was performed on full volumes using 3D sliding-window evaluation with 50\% overlap and test-time augmentation (horizontal flip). For ACT configurations, the same 128 $\times$ 128 $\times$ 64 crop geometry was used for both training and sliding-window inference. Experiments were implemented in PyTorch (v2.0.1)~\cite{paszke2019pytorch} and MONAI (v0.8.1)~\cite{cardoso2022monai}.

\subsection{Clustering-based appearance subtypes}
\label{sec:methods:clustering_subtypes}

To analyze appearance variability, we derived scan-level subdomains using unsupervised clustering. Seven features describing tumor intensity, boundary gradient, contrast, contrast-to-noise ratio, and tumor volume were extracted from each scan–segmentation pair. $k$-means clustering with feature standardization consistently favored $k=2$, and was fit on the training cohort and applied to the test cohort using the learned centroids (Section~\ref{appendix:clustering_details}, Figure~\ref{fig:sup_dataset_summary}). This resulted in two appearance subtypes: Cluster-B (high-contrast `bright' tumors) and Cluster-D (low-contrast `dark' tumors) that differed primarily in contrast and boundary definitions. In Figure~\ref{fig:tumor_aware_augmentation}, rows A–B show tumors from Cluster-B; rows C–D from Cluster-D, thereby illustrating that tumor-aware augmentation spans the observed intensity range across appearance subtypes.

\subsection{Feature similarity analysis}
\label{sec:methods:cka}

To characterize representational effects of tumor-aware augmentation, we computed linear centered kernel alignment (CKA)~\cite{kornblith2019similarity,cortes2012algorithms}. Backbone activations were extracted via forward hooks and adaptively pooled to 8 $\times$ 8 $\times$ 8 tokens prior to analysis. Similarity was estimated using minibatch-CKA with the unbiased HSIC estimator~\cite{nguyen2020wide}. We used linear CKA as a stable, established measure of layer-wise representation similarity; accordingly, these analyses should be interpreted as measuring linear feature reuse rather than nonlinear dependencies or tumor-region-weighted similarity.

\subsection{Evaluation metrics and statistical analysis}
\label{sec:methods:evaluation_protocols}

A structure was considered detected if its predicted segmentation overlapped the reference by $\geq 0.1$ Dice coefficient. Primary metrics were surface DSC (sDSC), volume ratio (VR), and detection rate with 95\% confidence intervals (Wilson score). sDSC at 2~mm tolerance was used as the primary overlap metric to align assessment with clinically-motivated tolerances where errors in the boundary are more problematic than the inside voxels that can often be addressed by basic post-processing~\cite{muller2022towards,nikolov2021clinically,vaassen2020evaluation}.

For missed cases, bounded penalties were assigned (sDSC = 0.1; VR = 10) so that missed detections remained worse than detected cases while preserving rank structure for paired testing; the main conclusions were unchanged when considering detection rate separately. Cases missed by all models were excluded from paired comparisons. Continuous metrics used the Wilcoxon signed-rank test with Bonferroni correction applied across all pairwise configuration comparisons within each metric; detection differences were reported in percentage points (pp) with McNemar test.

\section{Results}
\label{sec:results}

We denote the baseline cubic-input configuration as \textit{-Base}. Models using tumor-aware augmentation are appended with \textit{-TA}, and those combining anisotropic cropping with tumor-aware augmentation with \textit{-ACT}. We first report overall segmentation performance across configurations and then examine the two proposed failure modes: padding-induced token inefficiency using ADI and appearance mismatch using subtype-based analyses. Finally, we use CKA and CT-initialization versus Scratch comparisons to interpret how these effects relate to representation change and pretrained initialization.

\subsection{Segmentation performance}
\label{sec:results_seg_performance}

\begin{table}[t]
\def\arraystretch{1.2}
\centering
\caption{Segmentation performance on the rectal MRI test cohort ($N=247$) for the Base, tumor-aware augmentation (TA), and anisotropic crop (ACT) configurations across both backbones. Surface DSC (sDSC) and volume ratio (VR) are reported as median [IQR]; detection rate is reported with its Wilson 95\% confidence interval. sDSC and VR exclude cases missed by all three configurations within each backbone ($N=15$ for SMIT; $N=20$ for Swin~UNETR) and apply bounded penalties for individual misses (sDSC=0.1, VR=10). Detection rate is computed over all 247 scans.}
\label{tab:summary_results}
\resizebox{0.98\columnwidth}{!}{%
\begin{tabular}{lccc}
 & sDSC & VR & Det.\ rate \\
\midrule
SMIT-Base       & 0.559 [0.428--0.719] & 1.023 [0.622--2.003] & 86.2\% [81.4--90.0] \\
SMIT-TA         & 0.587 [0.426--0.749] & 1.001 [0.699--1.940] & 88.7\% [84.1--92.0] \\
SMIT-ACT        & 0.624 [0.462--0.763] & 1.046 [0.724--1.925] & 91.1\% [86.9--94.0] \\
\midrule
Swin~UNETR-Base & 0.532 [0.399--0.680] & 0.898 [0.523--1.889] & 78.5\% [73.0--83.2] \\
Swin~UNETR-TA   & 0.589 [0.445--0.736] & 0.991 [0.714--1.948] & 87.0\% [82.3--90.7] \\
Swin~UNETR-ACT  & 0.621 [0.461--0.752] & 1.036 [0.694--1.812] & 88.7\% [84.1--92.0] \\
\bottomrule
\end{tabular}%
}
\end{table}

Table~\ref{tab:summary_results} summarizes segmentation performance on the rectal MRI test cohort ($N=247$). Across both backbones, performance improved from the Base to TA to ACT configurations, indicating that both tumor-aware augmentation and anisotropic cropping contributed to segmentation robustness.

For the SMIT backbone, tumor-aware augmentation significantly improved the sDSC ($p=0.003$) and volume ratio ($p<0.001$) relative to the Base model, while detection increased by $+2.5$~pp ($p>0.05$). Anisotropic cropping resulted in a larger margin of improvement (sDSC and VR: $p<0.001$ versus Base), and detection reached 91.1\% ($+$ 4.9~pp versus Base). Volume ratio medians remained near unity across all configurations (0.898--1.046) despite broad per-case variability, indicating that the largest improvements were reflected more clearly in sDSC and detection rate. For Swin~UNETR, the Base configuration performed substantially weaker than the SMIT backbone. Tumor-aware augmentation alone produced large gains in sDSC, VR, and detection ($+$8.5~pp; $p<0.001$). Anisotropic cropping further improved segmentation performance, though the differences between TA and ACT were not significant after correction.

\begin{figure*}
    \centering
    \includegraphics[width=0.98\linewidth]{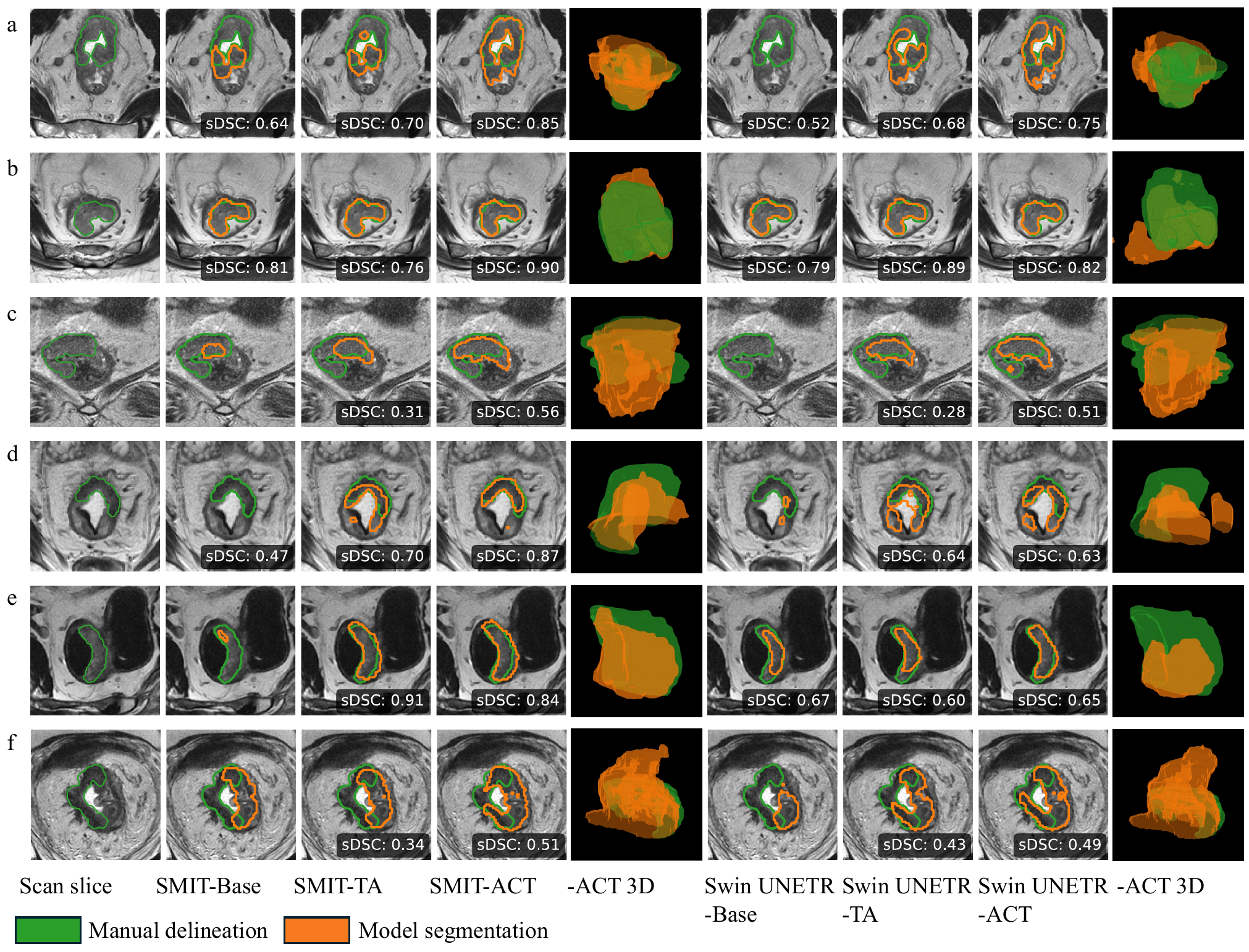}
    \caption{Rows A–C show representative high-contrast tumors (Cluster-B), while rows D–F show low-contrast tumors (Cluster-D). Each row presents the MRI slice with manual delineation (green), followed by predictions from SMIT and Swin~UNETR across the three configurations. Surface DSC (sDSC) is shown in the lower-right corner of each prediction panel. Missing values indicate that the tumor was not detected. The final column for each backbone shows representative 3D renderings of the -ACT segmentations.}
    \label{fig:segmentation_main_quali}
\end{figure*}

\begin{figure}
    \centering
    \includegraphics[width=0.98\linewidth]{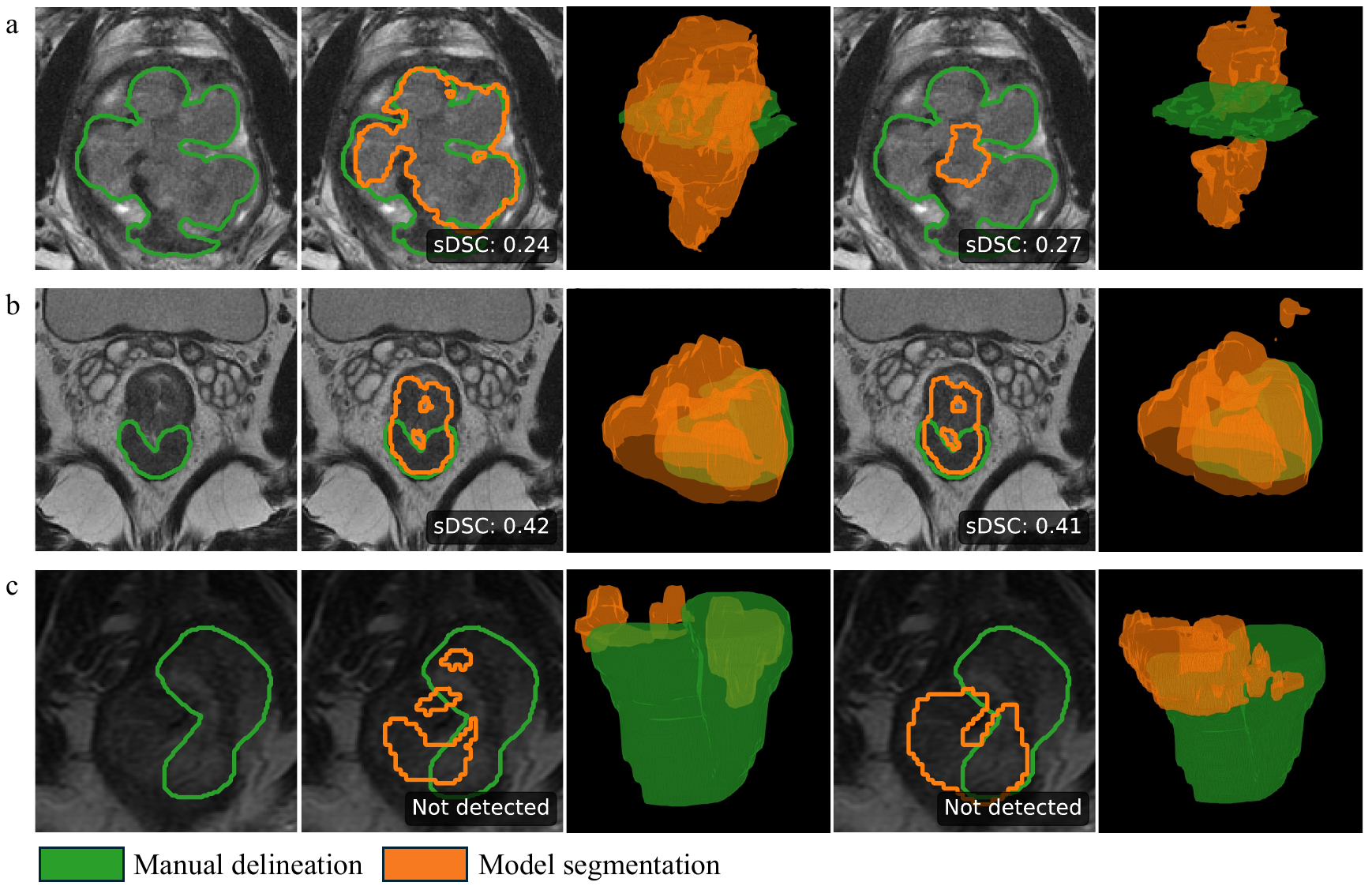}
    \caption{Representative challenging tumors under the ACT configuration. Each row shows an axial MRI slice with reference contour (green), followed by predictions from SMIT-ACT and Swin~UNETR-ACT (orange). Surface DSC (sDSC) is shown in the prediction panels; `Not detected' indicates failed tumor detection.}
\label{fig:act_limitations}
\end{figure}

Qualitative examples illustrate improved tumor delineation across configurations (Figure~\ref{fig:segmentation_main_quali}), while additional challenging cases under the final ACT configurations are shown in Figure~\ref{fig:act_limitations}. Together, these results indicate that both proposed interventions improve segmentation performance, but do not yet explain why. We therefore next examine the proposed token-efficiency failure mode using ADI.

\subsection{Mechanistic analyses of failure modes}
\label{sec:results_mechanistic}

\subsubsection{Zero-padding results in attention dilution}
\label{sec:results_padding_adi}

We next examined the proposed token-efficiency failure mode by quantifying the extent to which padding diverts attention away from real tokens. Among the 247 test cases, 193 (78.1\%) required zero-padding under the $128^3$ crop scheme used by the SMIT Base and TA configurations, with a median native $z$-extent of \iqr{93}{70}{120} voxels and a mean padding fraction ($pf$) of 0.270 $\pm$ 0.209. The padding fraction was negatively correlated with sDSC ($\rho$ = -0.21, $p=0.0015$). 

To quantify this relationship, we computed bootstrap confidence intervals for the association between Stage~3 ADI and segmentation performance in the SMIT intervention sequence. Under the cubic TA configuration, Stage~3 ADI was significantly negatively correlated with sDSC ($\rho$ = -0.172, 95\% CI: -0.298 to -0.042, $p=0.011$). After anisotropic cropping, the median padding fraction decreased to 0 and mean Stage~3 ADI decreased from 0.076 to 0.0067, while detection increased from 219/247 to 225/247 and median sDSC increased from 0.610 to 0.633. Because ACT largely removes padding-induced dilution, the residual ADI-sDSC association was weaker. This relationship was specific to continuous segmentation quality: Stage~3 ADI showed only a weak and inconsistent association with the binary detection endpoint (non-significant for SMIT, weakly significant for Swin~UNETR; Table~\ref{tab:adi_detection}), reinforcing that detection is best treated as a coarse thresholded endpoint rather than a direct ADI readout.

\begin{figure*}
    \centering
    \includegraphics[width=0.98\linewidth]{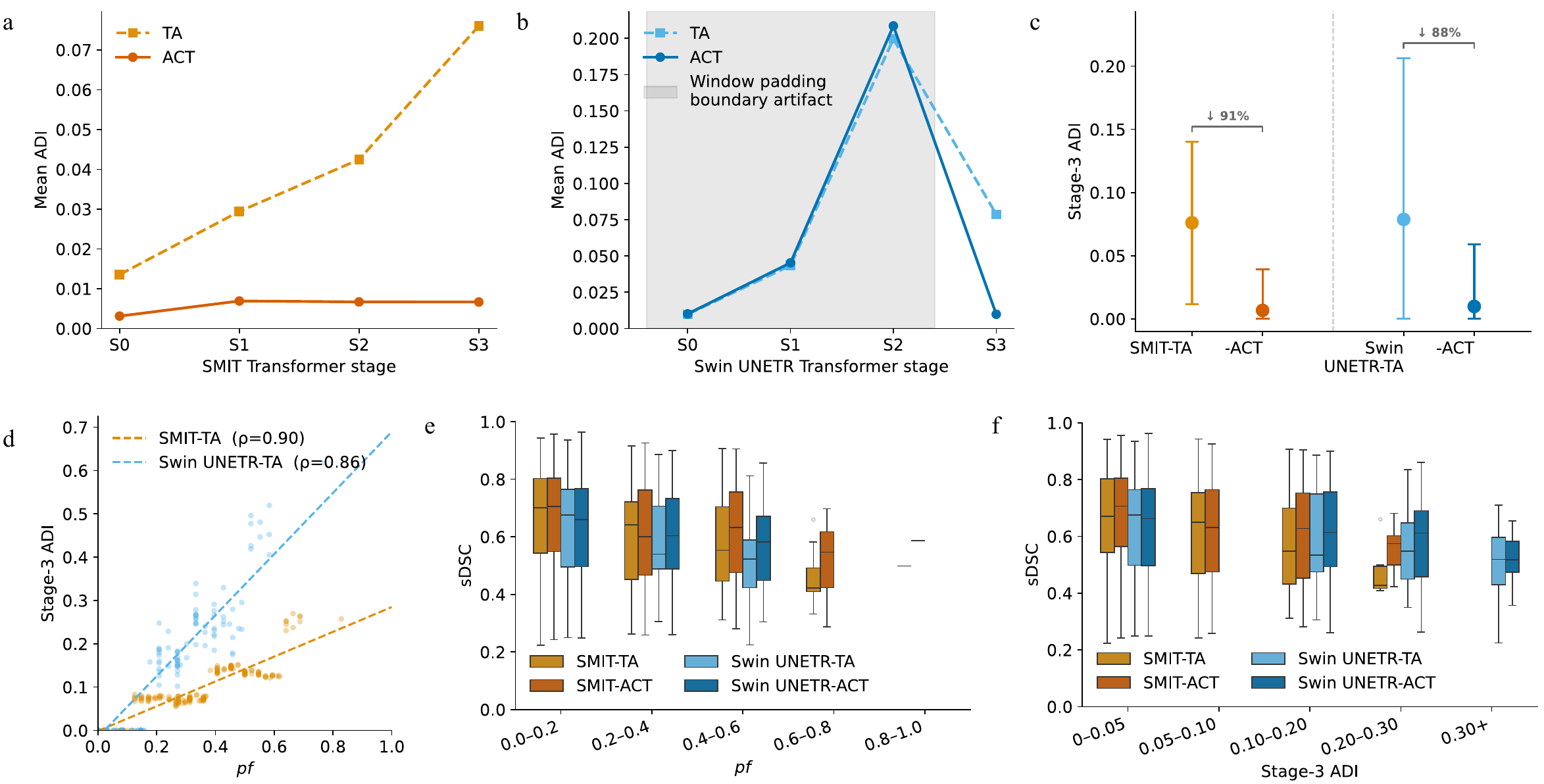}
    \caption{Attention dilution index (ADI) analysis linking hierarchical depth, padding fraction, and segmentation accuracy. (A–B) Mean ADI per stage for SMIT and Swin~UNETR; the shaded region highlights the S0–S2 window artifact in Swin~UNETR. (C) Mean Stage-3 ADI illustrating the effect of anisotropic cropping across both backbones. (D) Per-scan Stage-3 ADI as a function of padding fraction ($pf$), with per-backbone Spearman regression fits. (E–F) sDSC distributions binned by $pf$ and Stage-3 ADI.}
    \label{fig:dilution_results}
\end{figure*}

Under cubic inputs, ADI increased with hierarchical depth for both backbones (Figure~\ref{fig:dilution_results}A and B), indicating that padding-induced routing inefficiency was amplified through the hierarchical representations. This effect was substantially reduced by anisotropic cropping (Figure~\ref{fig:dilution_results}C). For SMIT, Stage~3 dilution reached 0.076 $\pm$ 0.064 and was strongly correlated with $pf$ ($\rho$ = 0.897, $p<0.0001$, Figure~\ref{fig:dilution_results}D) and negatively correlated with sDSC ($\rho$ = -0.172, $p=0.011$). For SMIT-ACT, Stage~3 dilution decreased to 0.007, a $\sim$91\% reduction relative to the cubic inputs. Across paired cases ($N=214$), anisotropic cropping improved sDSC by $+$0.025 ($p<0.001$) between SMIT-TA and SMIT-ACT. The largest gains occurred in heavily padded volumes ($pf \geq 0.5$, Figure~\ref{fig:dilution_results}E). Together, these results suggest that padding-induced attention dilution is associated with reduced segmentation accuracy (Figure~\ref{fig:dilution_results}F).

Repeating the analysis with Swin~UNETR produced the same qualitative pattern; $pf$ was also negatively correlated with sDSC ($\rho$ = -0.223, $p=0.001$). Stage~3 ADI remained strongly correlated with padding fraction ($\rho$ = 0.861, $p<0.001$, Figure~\ref{fig:dilution_results}D) and negatively correlated with sDSC ($\rho$ = -0.173, $p=0.011$), confirming that the dilution mechanism is not backbone-specific. However, Stage~0 to Stage~2 showed elevated dilution under ACT that was absent in SMIT (Figure~\ref{fig:dilution_results}B), which we attribute to the windowing artifact in Swin Transformers. Specifically, with $z$-crop 64, the intermediate feature-map extents along $z$ after patch merging (32, 16, and 8 voxels) are not all integer multiples of 7, causing the implementation to pad feature maps to the nearest multiple before window partitioning. This introduces additional zero tokens at window boundaries independently of MRI padding. The artifact is therefore a structural consequence of Swin~UNETR's larger window size; by contrast, SMIT's 4 $\times$ 4 $\times$ 4 windows divide evenly into the corresponding $z$-extents (16 and 8), so no boundary padding arises there. The artifact disappears at Stage~3, where the feature map collapses to a single window and the boundary effects vanish. Stage~3 dilution was reduced by $\sim$87\% under ACT ($0.079 \rightarrow 0.010$). Across paired cases ($N=210$), the improvement in sDSC between Swin~UNETR-TA and Swin~UNETR-ACT was small and non-significant ($+$ 0.005; $p=0.118$), consistent with the lower baseline padding severity under $96^3$ inputs. Overall, the findings for Swin~UNETR are consistent with the SMIT results (Figure~\ref{fig:dilution_results}E-F), confirming that the Stage~0-2 elevation reflects window arithmetic rather than a true increase in MRI-induced dilution. Overall, these results support token inefficiency induced by zero-padding as a genuine failure mode and explain why anisotropic cropping improved performance, particularly for SMIT and for heavily padded cases.

\subsubsection{Exploratory analysis of how tumor appearance heterogeneity impacts accuracy}
\label{sec:mixed_versus_clusters}

We performed unsupervised clustering of tumor appearance variations to examine the impact of appearance heterogeneity coverage on segmentation robustness. Specifically, the goal was to understand how tumor appearance variations (not to be confused with biological tumor heterogeneity) impacted accuracy.

\begin{figure}
    \centering
    \includegraphics[width=0.98\linewidth]{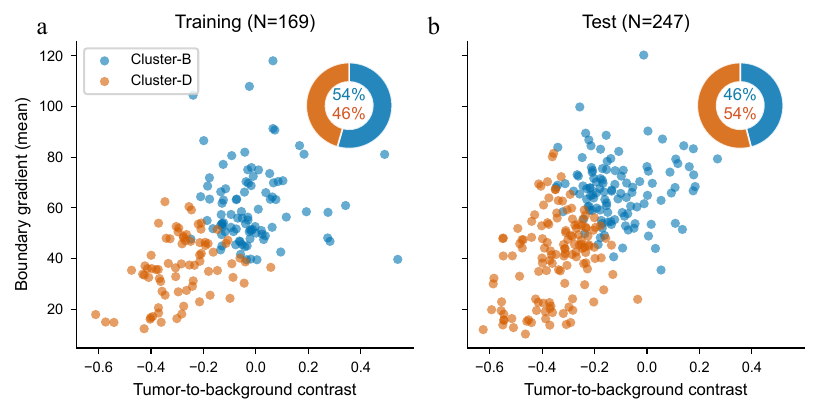}
    \caption{Two tumor subtypes identified by unsupervised clustering were consistent across cohorts. Tumor-to-background contrast versus boundary gradient for the training (A) and held-out test set (B). Cluster-B (blue) tumors show higher contrast and sharper boundaries than Cluster-D (orange). Donut insets show subtype proportions. Note that one contrast outlier was excluded from panel B.}
    \label{fig:sup_dataset_summary}
\end{figure}

Unsupervised clustering identified two reproducible appearance subtypes: Cluster-B, consisting of high-contrast `bright' tumors, and Cluster-D, consisting of lower-contrast `dark' tumors (Figure~\ref{fig:sup_dataset_summary}). The pretrained SMIT-ACT model achieved a median sDSC \iqr{0.690}{0.541}{0.786} with 96.5\% detection for Cluster-B and \iqr{0.522}{0.392}{0.717} with 85.7\% detection for Cluster-D, indicating that low-contrast tumors remained challenging but improved over the Base configuration.

Next, to study the importance of appearance coverage, we performed subtype-restricted training, wherein models were restricted to homogenized datasets with either dark or bright tumors. A model trained exclusively on Cluster-B retained high performance within Cluster-B (96.5\% detection; median sDSC 0.680) but collapsed to 58.6\% detection with median sDSC 0.506 on Cluster-D ($p<0.001$). In contrast, a model trained only on Cluster-D generalized more evenly across both subtypes, achieving 82.0\% detection with median sDSC 0.587 on Cluster-D and 87.7\% detection with median sDSC 0.594 on Cluster-B. These results suggest that segmentation robustness depends strongly on exposure to diverse tumor appearances and are consistent with the interpretation that the gains from tumor-aware augmentation arise, at least in part, from improved appearance coverage rather than generic regularization alone. This interpretation is also consistent with the design of the tumor-aware augmentation (Figure~\ref{fig:tumor_aware_augmentation}), which broadens local tumor intensity patterns without perturbing background anatomy. 

This same pattern was reflected in the residual failure cases under the final ACT configurations. For SMIT-ACT, 19 of 23 missed tumors belonged to Cluster-D, while for Swin~UNETR-ACT, 22 of 28 misses were in Cluster-D. Missed tumors were also smaller and less conspicuous than detected tumors: under SMIT-ACT, the median tumor volume was 3,405 mm$^3$ for missed versus 13,113 mm$^3$ for detected cases, with lower median tumor-to-background contrast (-0.364 versus -0.229) and lower boundary gradient (42.4 versus 54.8). A similar pattern was observed for Swin~UNETR-ACT, indicating that the principal residual blind spots were small, low-contrast tumors with poorly defined boundaries.

\subsection{Feature similarity analysis}
\label{sec:results_cka}

We next investigated how the proposed interventions altered learned representations. To this end, we computed layer-wise feature similarity between pretrained and fine-tuned models using linear CKA.

\begin{figure*}[t]
    \centering
    \includegraphics[width=0.98\linewidth]{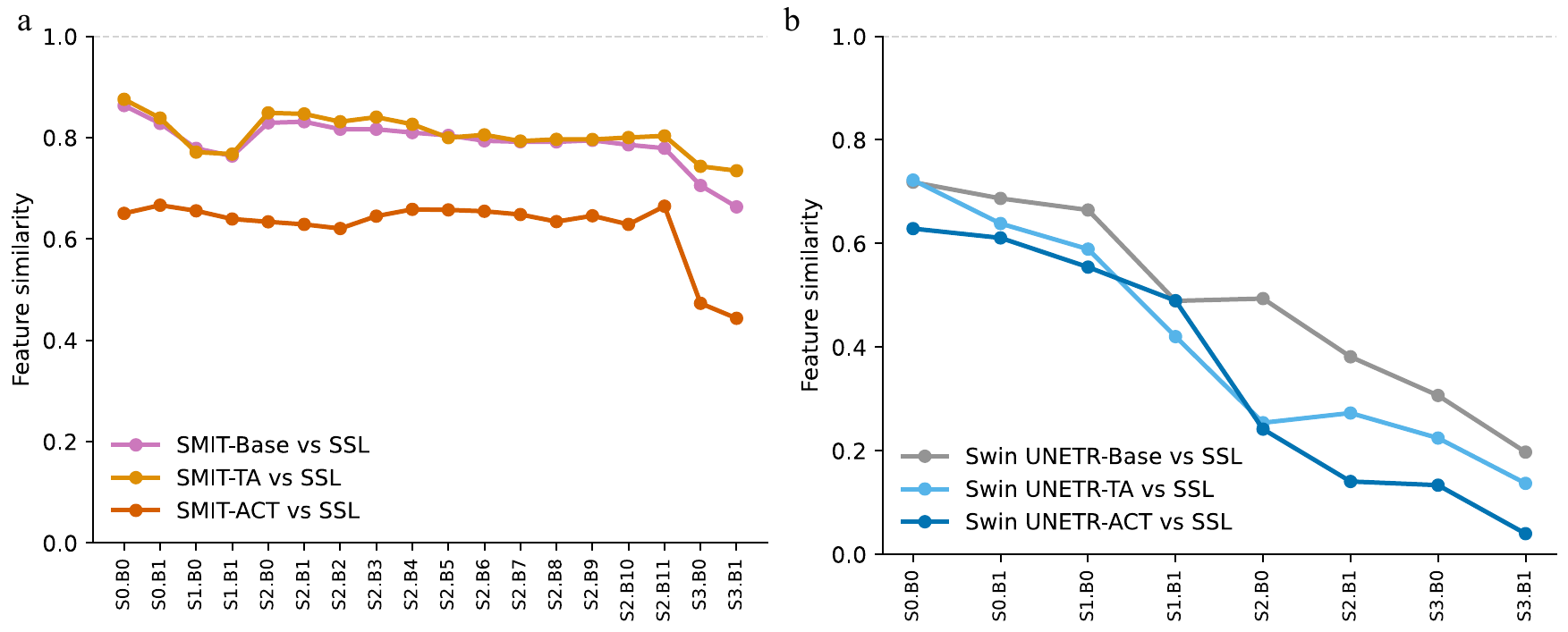}
    \caption{Layer-wise feature similarity to the pretrained backbone measured using linear CKA. Diagonal CKA across transformer blocks for (a) SMIT and (b) Swin~UNETR under different configurations.}
    \label{fig:cka_group_models_vs_ssl}
\end{figure*}

For both backbones, early layers remained relatively similar to the pretrained representations, while deeper layers progressively diverged following fine-tuning (Figure~\ref{fig:cka_group_models_vs_ssl}). Among the fine-tuned configurations, ACT produced the largest representational shift, reflected by the least feature reuse in the deepest stages, consistent with the stronger performance gains observed under anisotropic cropping. In contrast, the Base and TA configurations remained closer to the pretrained representations across most layers, indicating that tumor-aware augmentation primarily influenced later-stage feature refinement rather than the earliest feature extractors. Swin~UNETR also exhibited broader layer-wise drift than SMIT (Figures~\ref{fig:cka_smit_adv} and~\ref{fig:cka_swin_unetr_adv}).

\begin{figure}
    \centering
    \includegraphics[width=0.98\linewidth]{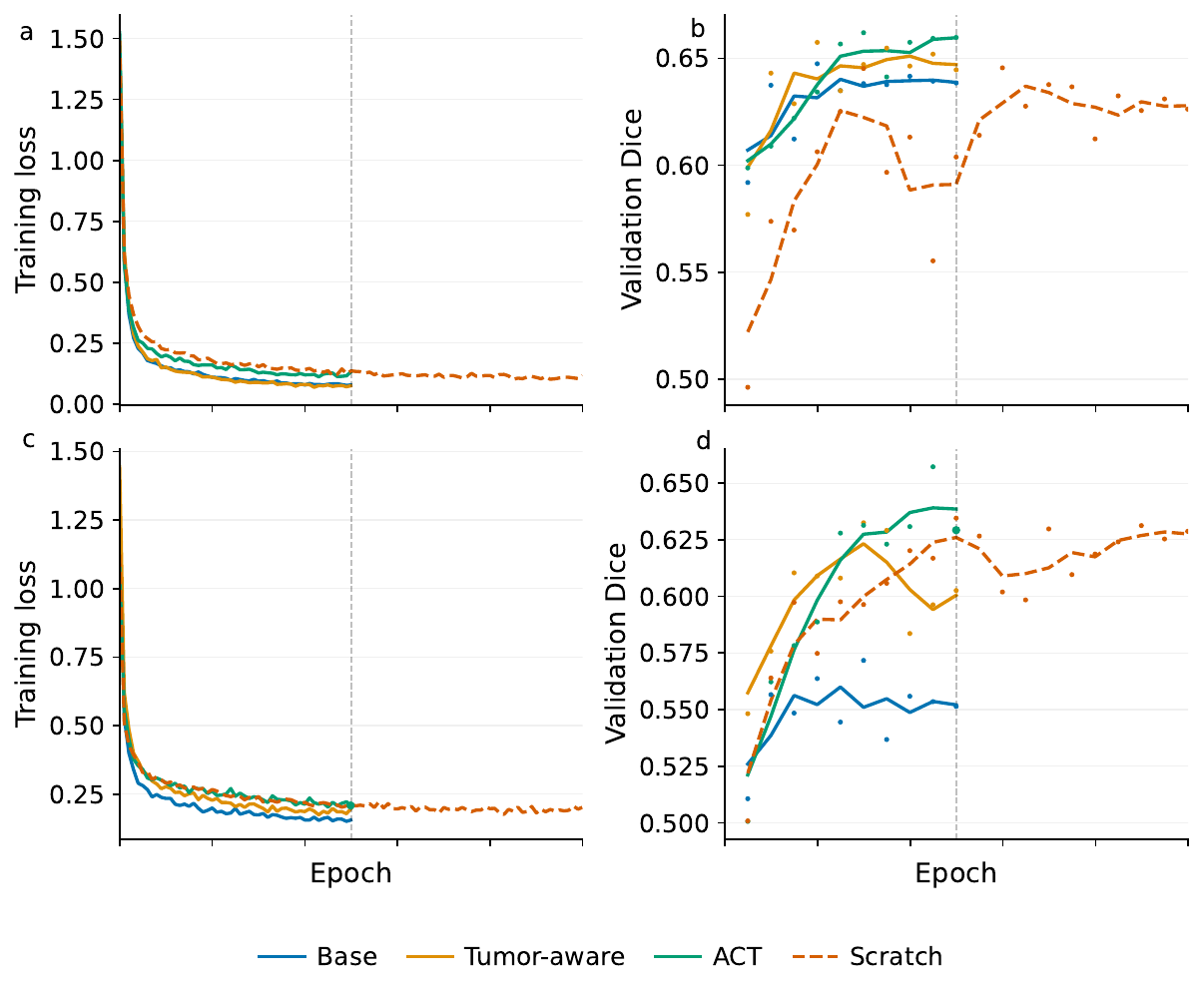}
    \caption{Training convergence during rectal MRI fine-tuning. (a) SMIT training loss. (b) SMIT validation Dice. (c) Swin~UNETR training loss. (d) Swin~UNETR validation Dice. Colors denote Base, tumor-aware, ACT, and Scratch configurations; points show raw validation measurements and lines show lightly smoothed trends. The dashed line marks epoch 500, the reported training horizon for the pretrained configurations, while the randomly initialized (Scratch) was trained for twice the epochs.}
    \label{fig:sup_training_convergence}
\end{figure}

Taken together, these results show that greater feature reuse relative to the pretrained backbone does not necessarily predict better downstream performance. Instead, the strongest-performing ACT configurations exhibited the largest representational shift in the deeper layers, suggesting that effective transfer in this setting depends more on task-specific adaptation than on preserving pretrained features. To ensure that differences in performance were not attributable to under-training, we examined training loss and validation Dice trajectories for both backbones across configurations. Validation performance largely plateaued by the reported training horizon for the pretrained models (Figure~\ref{fig:sup_training_convergence}).

\subsection{Pretraining versus random initialization}
\label{sec:ct_versus_random_init}

Having analyzed the two task-specific failure modes, we finally asked what benefit remained from pretraining itself once padding burden was controlled. We therefore compared pretrained and randomly initialized (Scratch) models under the fixed ACT configuration (Table~\ref{tab:pretraining_ablation}).

\begin{table}[t]
\def\arraystretch{1.2}
\centering
\caption{Effect of CT initialization under the ACT configuration for SMIT and Swin~UNETR. Surface DSC (sDSC) and volume ratio (VR) are reported as median [IQR]; detection rate is reported with its Wilson 95\% confidence interval. To enable paired comparison between pretrained and Scratch models, sDSC and VR exclude cases missed by both configurations within each backbone ($N=21$ for SMIT; $N=24$ for Swin~UNETR), and apply bounded penalties for individual misses (sDSC=0.1, VR=10). Because this exclusion set differs from that used in Table~\ref{tab:summary_results} (which excludes cases missed by all three configurations), the IQR for the pretrained models may differ slightly between the two tables. Detection rate is computed over all 247 scans.}
\label{tab:pretraining_ablation}
\resizebox{0.98\columnwidth}{!}{%
\begin{tabular}{lccc}
Model & sDSC & VR & Det.\ rate \\
\midrule
\multicolumn{4}{l}{\textit{SMIT-ACT}} \\
Pretrained & 0.624 [0.462--0.763] & 1.025 [0.714--1.840] & 91.1\% [86.9--94.0] \\
Scratch    & 0.618 [0.441--0.750] & 0.965 [0.630--1.822] & 88.3\% [83.6--91.7] \\
\midrule
\multicolumn{4}{l}{\textit{Swin~UNETR-ACT}} \\
Pretrained & 0.621 [0.461--0.752] & 1.021 [0.693--1.769] & 88.7\% [84.1--92.0] \\
Scratch    & 0.624 [0.458--0.751] & 0.972 [0.679--1.694] & 88.3\% [83.6--91.7] \\
\bottomrule
\end{tabular}%
}
\end{table}

For the SMIT backbone, CT initialization significantly improved sDSC ($p<0.001$), while detection increased by $+$2.4~pp ($p>0.05$), although the latter difference was not significant. We note that although CT pretraining significantly improved sDSC for SMIT, the absolute improvement was modest compared with the larger gains achieved through tumor-aware augmentation and anisotropic cropping. This suggests that pretraining provided a residual benefit, but was not the dominant factor driving segmentation performance. Volume ratio did not show a consistent directional improvement despite a nominally significant Wilcoxon result, suggesting distributional skew rather than a robust effect. In contrast, Swin~UNETR showed no significant difference between pretraining and scratch-trained models across any metrics (for example, sDSC: $p=0.110$).

To further isolate the effect of pretraining source, we also evaluated Swin~UNETR initialized with the in-house pretrained weights (the same 10,432 CT corpus used for SMIT) under the ACT configuration with tumor-aware augmentation. In this supplementary comparison, the test-cohort mean sDSC was 0.596 $\pm$ 0.174, numerically lower than both the NVIDIA-pretrained (0.621) and randomly initialized (0.624) Swin~UNETR-ACT models. This result further supports the interpretation that, in the evaluated settings, CT-based pretraining source and initialization provided utmost limited additional advantage for rectal MRI tumor segmentation once the task-specific failure modes were addressed by tumor-aware augmentation and anisotropic cropping.

\begin{figure}
    \centering
    \includegraphics[width=0.98\linewidth]{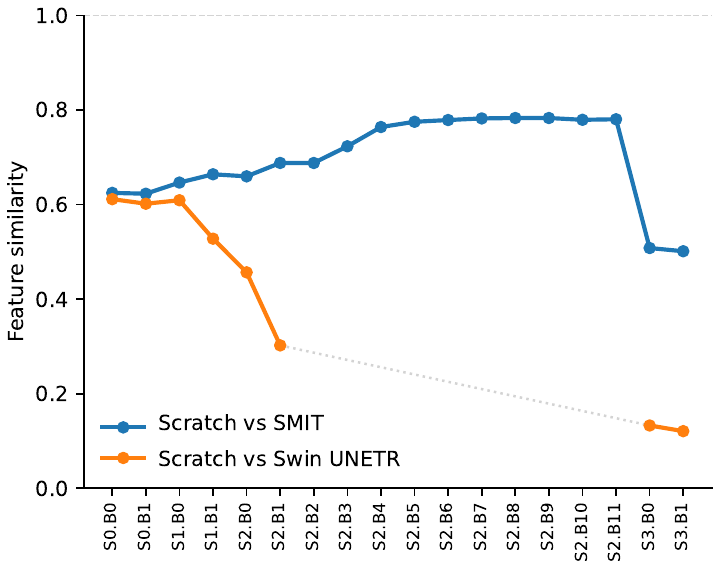}
    \caption{Layer-wise feature similarity between randomly initialized (Scratch) and CT-pretrained (CT-Init.) models measured using diagonal linear CKA across transformer blocks for SMIT and Swin~UNETR.}
    \label{fig:cka_group_models_ssl_vs_random}
\end{figure}

These results indicate that once the padding burden is reduced, pretraining provides a modest benefit for SMIT but does not substantially affect Swin~UNETR. In both cases, the dominant gains observed earlier arise from tumor-aware augmentation and reduced padding rather than from weight initialization alone. Consistent with this, ADI was indistinguishable between pretrained and Scratch models, and CKA showed only limited layer-wise divergence between them (Figures~\ref{fig:cka_group_models_ssl_vs_random} and~\ref{fig:cka_ct_init_vs_random}). Together, these findings suggest that CT pretraining offered at most a modest and architecture-dependent additional benefit without task-specific corrections for appearance mismatch and token inefficiency.

\subsection{Generalization to large-scale pretraining}
\label{subsec:additional_baselines}

As a supporting benchmark, we compared the final SMIT-ACT configuration with VoCo-ACT~\cite{wu2025large}; this comparison was included as a supporting benchmark rather than as a mechanistic ADI analysis, and VoCo differs from the primary Swin~UNETR baseline in architecture version (it uses Swin-V2~\cite{he2023swinunetr} instead of Swin-V1~\cite{hatamizadeh2021swin}), pretraining setup (volume contrast instead of image inpainting), and number of pretraining examples (5,050 versus $\geq$ 160,000 CT scans). After applying the same tumor-aware augmentation and anisotropic cropping strategy, VoCo-ACT was more accurate and achieved a higher detection rate than VoCo-Base, confirming our findings with transformer models pretrained with substantially smaller pretraining datasets.

\begin{table}[t]
\centering
\def\arraystretch{1.2}
\caption{Additional cross-baseline comparison on the held-out test cohort. sDSC and VR were summarized after excluding cases missed by both VoCo configurations, with bounded penalties applied to individual missed detections within the remaining cases. Detection rate was computed over all 247 test scans.}
\label{tab:additional_baselines}
\resizebox{0.98\columnwidth}{!}{%
\begin{tabular}{lccc}
Model & sDSC & VR & Det.\ rate \\
\midrule
VoCo-Base & 0.599 [0.485--0.746] & 0.991 [0.724--1.899] & 88.0\% \\
VoCo-ACT & 0.628 [0.438--0.746] & 1.000 [0.744--1.751] & 90.3\% \\
\bottomrule
\end{tabular}
}
\end{table}

\section{Discussion and conclusion}
\label{sec:discussion_and_conclusion}

A key finding of this study was that CT-pretrained models fail to transfer to MRI tumor segmentation tasks under two interacting failure modes resulting from padding-induced attention dilution and insufficient appearance coverage. Although our results showed pretrained representations were beneficial compared to scratch-trained models as shown in prior works~\cite{jiang2022self,tang2022self,wang2025triad}, the benefit was marginal compared to prior studies performing same-modality transfer for tumor segmentation~\cite{chen2019med3d,jiang2025self}. These findings support a more nuanced interpretation of transfer learning in medical imaging: while pretraining may provide useful initialization and improve optimization efficiency, it may not fully compensate for substantial geometric and statistical discrepancies between pretraining and downstream domains. Our CKA analyses further support this interpretation, showing that early layers remained relatively stable while larger representational shifts occurred in deeper stages associated with task-specific boundary resolution~\cite{kornblith2019similarity,raghu2019transfusion,matsoukas2022makes}.

Our proposed tumor-aware augmentation and anisotropic cropping strategies mitigated these issues and resulted in improved accuracy despite substantial modality shift. Different from prior studies primarily describing negative modality transfer, our work systematically investigated why these transfer failures occur and demonstrated that task-specific interventions improved robustness, with detection increasing by 4.9~pp for SMIT and 10.2~pp for Swin~UNETR from Base to ACT. Tumor-aware augmentation improved performance across heterogeneous tumor appearances, while anisotropic cropping reduced padding burden and computational inefficiency. Although low-contrast tumors with diffuse boundaries remained challenging, the proposed framework improved detection performance and reduced the number of complete misses in these difficult cases.

ADI is therefore intended as a padding-aware diagnostic rather than a general-purpose attention interpretability metric. Existing attention analyses such as attention rollout and attention flow estimate how attention propagates through transformer layers~\cite{abnar2020quantifying}, while representation analyses of vision transformers characterize layer-wise feature structure, spatial localization, and attention behavior~\cite{raghu2021vision}. In contrast, standard attention entropy measures dispersion of attention weights without distinguishing whether this dispersion is assigned to informative anatomical tokens or non-informative padding tokens. ADI addresses this distinction by conditioning on real anatomical query tokens and decomposing their attention entropy by key-token type, and quantifies padding-induced token inefficiency caused by input-size mismatch, rather than attributing individual predictions to image regions.

This study has a few limitations. First, the experiments were restricted to CT-to-MRI transfer for rectal cancer segmentation using pretrained hierarchical transformer architectures, and whether these failure modes generalize to other organs, modalities, and architectures remains to be established. Accordingly, our findings should be interpreted as evidence that cross-modal pretraining can fail under specific out-of-distribution conditions involving appearance mismatch and geometric incompatibility, rather than as evidence of a universal limitation of transfer learning. Additionally, evaluation was performed using retrospectively collected data from a single institution; although the development cohort included 1.5T and 3T GE Healthcare systems, the independent held-out test cohort was acquired primarily on a 1.5T GE Signa HDx scanner, limiting assessment of robustness across vendors, field strengths, acquisition protocols, institutions, and treatment paradigms. Future multi-center studies are needed to evaluate generalizability under broader clinical variability. Nevertheless, this study provides an early mechanistic investigation of transferability failures in the challenging setting of rectal MRI tumor segmentation. Future work should compare anisotropic cropping against masking-based approaches that preserve full field-of-view while explicitly excluding padding tokens from attention computation, and should also compare these task-specific interventions with conventional domain adaptation strategies, including adversarial feature alignment, image translation, and pseudo-labeling frameworks.

To support reproducibility, we will release the code for model training, inference, evaluation, and ADI computation upon publication, together with the configurations and preprocessing pipelines used in this study. Although the dataset cannot be publicly released, these resources will facilitate independent evaluation and extension of our framework. Overall, this work provides new insights into why cross-modal transfer learning can fail for transformer-based tumor segmentation and demonstrates practical strategies to improve robustness under substantial modality shift.

\appendix
\renewcommand{\thefigure}{A.\arabic{figure}}
\renewcommand{\thetable}{A.\arabic{table}}
\setcounter{figure}{0}
\setcounter{table}{0}

\section{Clustering pipeline details}
\label{appendix:clustering_details}

\begin{figure*}
    \centering
    \includegraphics[width=0.98\linewidth]{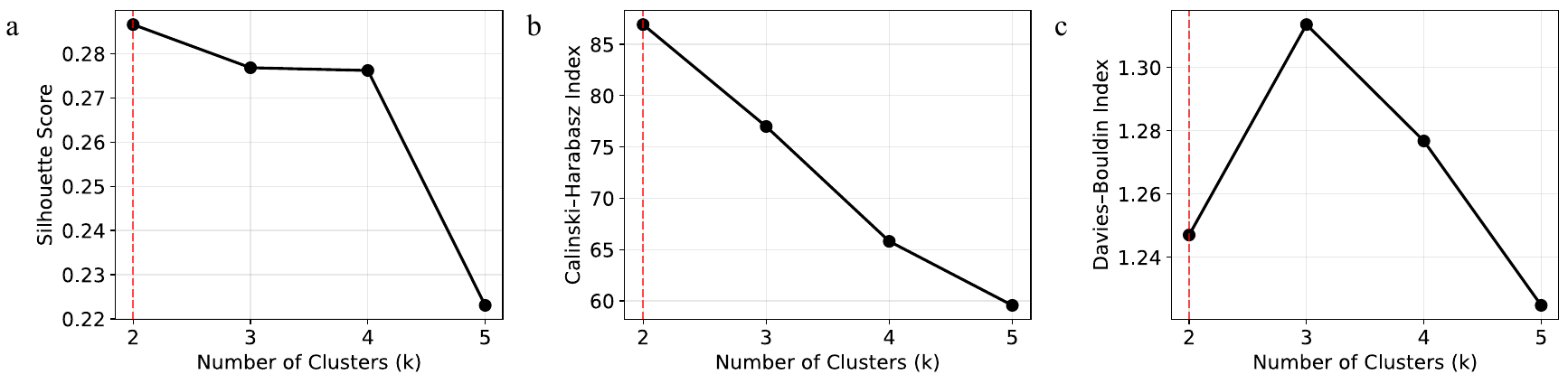}
    \caption{Cluster validity metrics across $k \in [2,5]$ for tumor subtype clustering. The dashed red line indicates the selected $k=2$ for (a) silhouette score, (b) Calinski–Harabasz index, and (c) Davies–Bouldin index. All three metrics converged on $k=2$ as optimal.}
    \label{fig:sup_tumor_clustering}
\end{figure*}

\begin{figure*}
    \centering
    \includegraphics[width=0.98\linewidth]{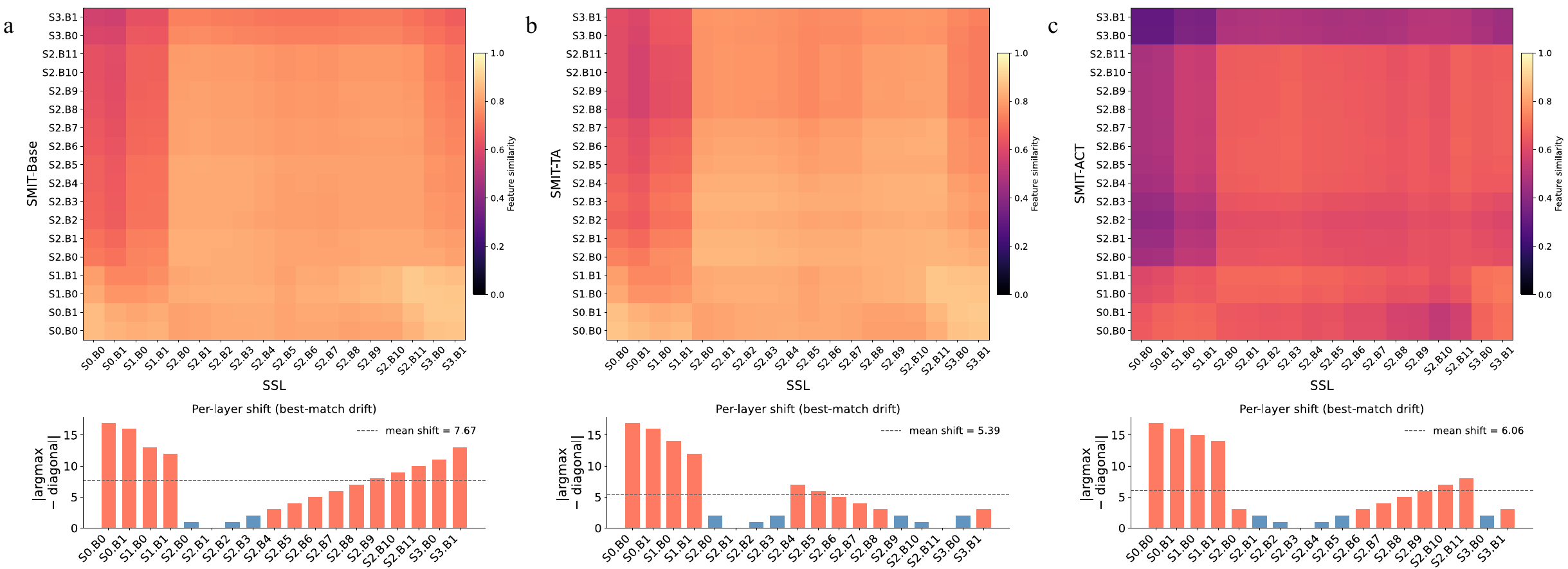}
    \caption{Feature similarity between the pretrained SMIT backbone and finetuned configurations. Linear CKA heatmaps compare the pretrained SSL model with (A) SMIT-Base, (B) SMIT-TA, and (C) SMIT-ACT. Bar plots summarize per-layer drift from the diagonal alignment that quantifies the degree of representational change; larger values indicate greater divergence.}
    \label{fig:cka_smit_adv}
\end{figure*}

Background intensities were computed from all non-tumor voxels with non-zero intensity to avoid padding effects. Boundary voxels were defined as the three-voxel tumor boundary, and boundary gradient was computed as the mean gradient magnitude at this boundary.

Features were standardized using z-scoring (zero mean, unit variance) prior to clustering to prevent scale-dominant variables from disproportionately influencing the cluster assignments. K-means clustering was fit only on the training split; the fitted standardization transform and cluster centroids were then applied to the test cohort to assign subtypes using the same feature definitions. 

The number of clusters was evaluated across $k \in [2,5]$ using silhouette score, Calinski–Harabasz index, and Davies–Bouldin index; all metrics converged on $k=2$ (Figure~\ref{fig:sup_tumor_clustering}). The resulting clusters corresponded to high-contrast bright well-defined tumors (Cluster-B) and low-contrast, dark, poorly visible tumors (Cluster-D) (Figure~\ref{fig:sup_dataset_summary}).

To assess stability, we repeated the clustering on 1,000 bootstrap resamples of the training cohort and compared the resulting assignments with the original $k=2$ solution after label matching. The median assignment agreement was 0.96 (95\% bootstrap interval: 0.83--1.00), supporting the use of the two-cluster solution as a stable exploratory appearance stratification.

\section{Understanding Centered Kernel Alignment (CKA)}
\label{appendix:cka}

We measured similarity between feature representations using centered kernel alignment (CKA)~\cite{kornblith2019similarity,cortes2012algorithms}.
Given two activation matrices $\bm{M}, \bm{N} \in \mathbb{R}^{n \times d}$ containing features from $n$ samples, CKA computes the normalized similarity between their Gram matrices using the Hilbert–Schmidt Independence Criterion (HSIC):

\begin{equation}
\mathrm{CKA}(\bm{M},\bm{N}) =
\frac{\mathrm{HSIC}_0(\bm{K},\bm{L})}
{\sqrt{\mathrm{HSIC}_0(\bm{K},\bm{K}),
\mathrm{HSIC}_0(\bm{L},\bm{L})}}
\label{eqn:CKA1}
\end{equation}

where $\bm{K} = \bm{M}\bm{M}^\top$ and $\bm{L} = \bm{N}\bm{N}^\top$ are the corresponding Gram matrices.

Computing CKA requires storing activations for the entire dataset, which is impractical for large transformer models.
Therefore, we employed minibatch CKA~\cite{nguyen2020wide}, obtained by averaging HSIC estimates across $k$ minibatches:

\begin{equation}
\mathrm{CKA}_{\mathrm{mb}} = \frac{A}{\sqrt{B \cdot C}}\,,
\end{equation}
\begin{gather*}
A = \tfrac{1}{k}{\textstyle\sum_{i=1}^{k}}\,\mathrm{HSIC}_1(\bm{K}_i,\bm{L}_i)\,,\\[4pt]
B = \tfrac{1}{k}{\textstyle\sum_{i=1}^{k}}\,\mathrm{HSIC}_1(\bm{K}_i,\bm{K}_i)\,,\\[4pt]
C = \tfrac{1}{k}{\textstyle\sum_{i=1}^{k}}\,\mathrm{HSIC}_1(\bm{L}_i,\bm{L}_i)\,,
\end{gather*}

where $\bm{K}_i = \bm{M}_i\bm{M}_i^\top$ and $\bm{L}_i = \bm{N}_i\bm{N}_i^\top$ are Gram matrices computed from the $i^{\text{th}}$ minibatch.

To reduce the dependence of CKA on batch size, we used the unbiased estimator of HSIC~\cite{song2012feature}:

\begin{multline}
\mathrm{HSIC}_1(\bm{K},\bm{L}) =
\frac{1}{n(n-3)}
\biggl[
\mathrm{tr}(\tilde{\bm{K}}\tilde{\bm{L}})\\
+\frac{\bm{1}^\top \tilde{\bm{K}}\bm{1}\;\cdot\;
\bm{1}^\top \tilde{\bm{L}}\bm{1}}{(n-1)(n-2)}
-\frac{2}{n-2}\,\bm{1}^\top \tilde{\bm{K}}\tilde{\bm{L}}\bm{1}
\biggr]
\end{multline}

where $\tilde{\bm{K}}$ and $\tilde{\bm{L}}$ are obtained by zeroing the diagonal entries of $\bm{K}$ and $\bm{L}$.

\begin{figure*}
    \centering
    \includegraphics[width=0.98\linewidth]{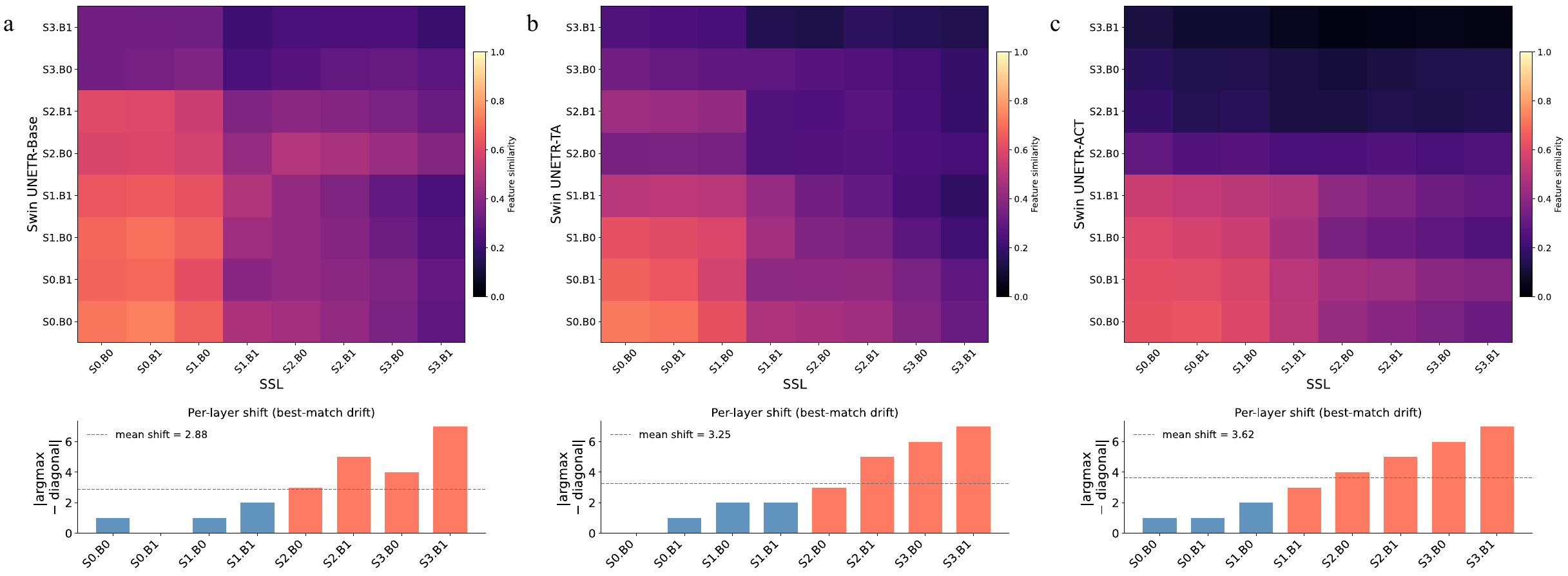}
    \caption{Feature similarity between the pretrained Swin~UNETR backbone and finetuned configurations. Linear CKA heatmaps compare the pretrained SSL model with (A) Swin~UNETR-Base, (B) Swin~UNETR-TA, and (C) Swin~UNETR-ACT. Bar plots summarize per-layer drift from the diagonal alignment that quantifies the degree of representational change; larger values indicate greater divergence.}
    \label{fig:cka_swin_unetr_adv}
\end{figure*}

\begin{figure*}
    \centering
    \includegraphics[width=0.98\linewidth]{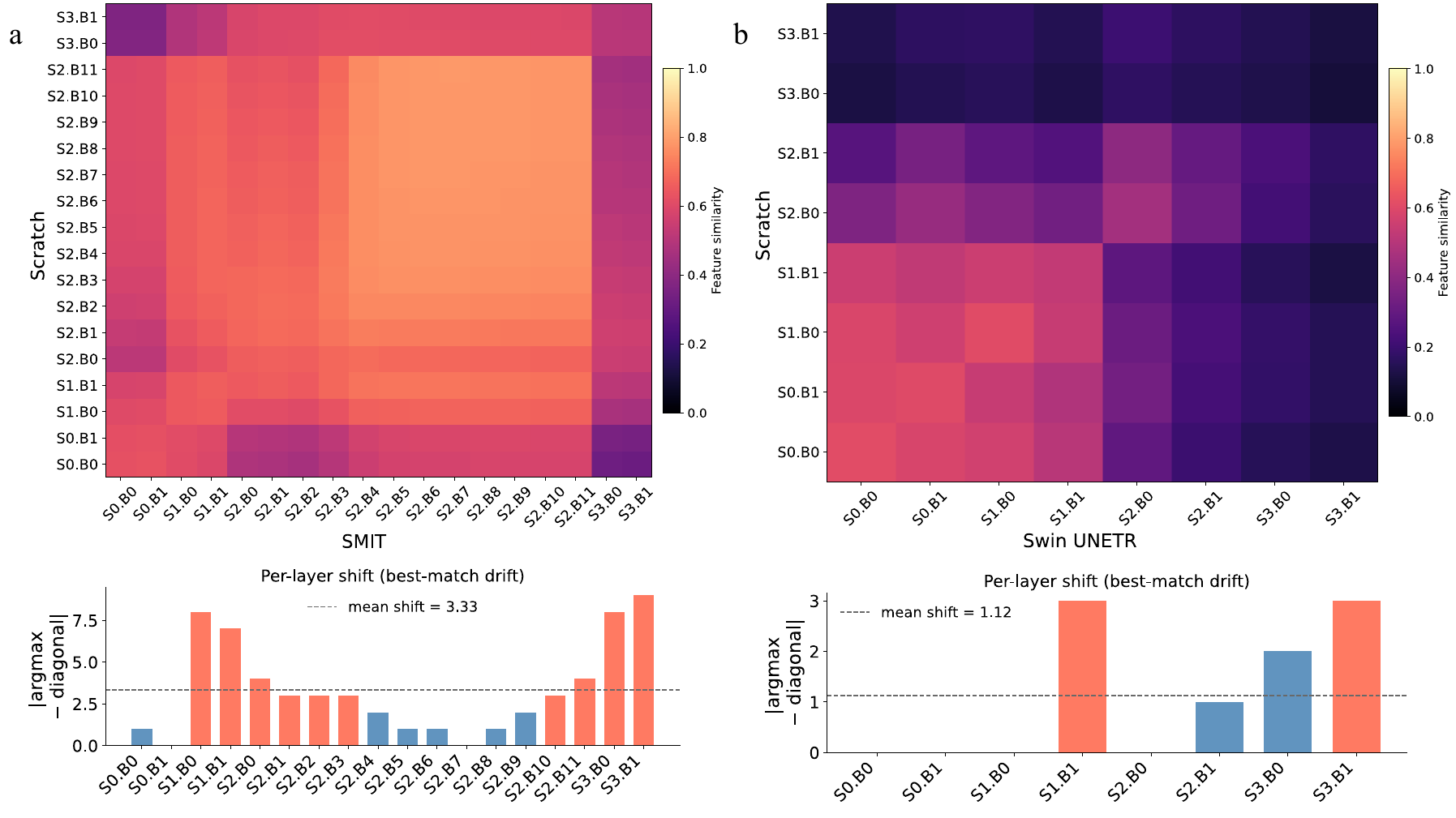}
    \caption{Feature similarity between CT-initialised and randomly initialized models. Linear CKA heatmaps compare layer-wise representations between pretrained and Scratch models for (A) SMIT and (B) Swin~UNETR. Bar plots summarize per-layer drift from the diagonal alignment that quantifies the degree of representational change; larger values indicate greater divergence.}
    \label{fig:cka_ct_init_vs_random}
\end{figure*}

\section{Supplementary methodological analyses}
\label{appendix:additional_results}

\subsection{Crop-depth padding sensitivity} \label{subsec:crop_depth}

To make the anisotropic crop-depth choice quantitative rather than purely architectural, we computed the through-plane zero-padding burden of the held-out cohort across candidate depths (Table~\ref{tab:crop_depth}). Padding fraction $pf=\max(0,\,C-L)/C$ for axial length $L$ and crop depth $C$. The burden rises steeply and monotonically with depth: 64 slices leaves only 13\% of volumes padded (median $pf=0$), whereas the cubic 128 depth pads 78\% of volumes (median $pf=0.27$). Depth 64 is thus the lowest architecture-compatible depth that keeps the median volume padding-free while halving token cost.

\begin{table}[t]
\centering
\def\arraystretch{1.2}
\caption{Through-plane zero-padding burden of the held-out test cohort ($N=247$, 1\,mm isotropic, median axial extent 93 voxels) as a function of crop depth. The right column is the relative token / windowed-attention cost versus the cubic 128 depth.}
\label{tab:crop_depth}
\resizebox{0.98\columnwidth}{!}{%
\begin{tabular}{lcccc}
Crop depth & Padding & mean $pf$ & median $pf$ & rel.\ tokens \\
\midrule
48  & 4.0\%  & 0.006 & 0.000 & 38\% \\
\textbf{64}  & \textbf{13.4\%} & \textbf{0.027} & \textbf{0.000} & \textbf{50\%} \\
80  & 32.8\% & 0.072 & 0.000 & 62\% \\
96  & 56.3\% & 0.134 & 0.031 & 75\% \\
112 & 67.6\% & 0.203 & 0.170 & 88\% \\
128 & 78.1\% & 0.270 & 0.273 & 100\% \\
\bottomrule
\end{tabular}
}
\end{table}

\subsection{Computational footprint of the interventions} \label{subsec:compute}

Because windowed-attention cost scales with token count (crop volume), anisotropic cropping \emph{reduces} the dominant transformer compute relative to cubic inputs (Table~\ref{tab:compute}). For SMIT, the 128 $\times$ 128 $\times$ 64 crop is 50\% of the cubic 128 $\times$ 128 $\times$ 128 token/attention cost, consistent with the measured 56\% wall-clock reduction (the remainder arising from non-attention layers and I/O). Tumor-aware augmentation is a train-time CPU-side mask intensity transform with no added parameters and no inference-time overhead.

\begin{table}[t] 
\centering
\def\arraystretch{1.2}
\caption{Computational footprint of the proposed interventions. Windowed-attention cost scales with token count. Anisotropic cropping reduces compute relative to cubic inputs; tumor-aware augmentation adds only a train-time CPU intensity transform with no added parameters and no inference-time cost.}
\label{tab:compute}
\resizebox{0.98\columnwidth}{!}{%
\begin{tabular}{lccc}
Configuration & Crop & Voxels & Rel. attn. FLOPs \\
\midrule
SMIT -Base/TA       & 128 $\times$ 128 $\times$ 128 & 2,097,152   & 100\% \\
SMIT -ACT           & 128 $\times$ 128 $\times$ 64  & 1,048,576   & 50\% \\
Swin~UNETR -Base/TA & 96 $\times$ 96 $\times$ 96    & 884,736     & 42\% \\
Swin~UNETR -ACT     & 96 $\times$ 96 $\times$ 64    & 589,824     & 28\% \\
\bottomrule
\end{tabular}
}
\end{table}

\subsection{Association between ADI and detection} 
\label{subsec:adi_detection} 

We separated the continuous and binary segmentation endpoints when relating Stage-3 ADI to performance (Table~\ref{tab:adi_detection}). ADI was significantly negatively correlated with \emph{continuous} sDSC for both backbones, but its association with the \emph{binary} detection endpoint was weak and inconsistent;  non-significant for SMIT and only weakly significant for Swin~UNETR. This supports interpreting ADI as a diagnostic of continuous segmentation quality and token efficiency, while treating detection as a coarser thresholded endpoint.

\begin{table}[t]
\centering
\def\arraystretch{1.2}
\caption{Association between Stage-3 ADI and segmentation performance under the cubic (\textit{-TA}) configuration. 95\% bootstrap CIs (5,000 resamples). $r_{pb}$: point-biserial correlation between ADI and detection rate; MW: Mann-Whitney test of higher ADI in missed cases.}
\label{tab:adi_detection}
\resizebox{0.98\columnwidth}{!}{%
\begin{tabular}{lcccc}
Backbone & \multicolumn{2}{c}{ADI vs sDSC} & \multicolumn{2}{c}{ADI vs Det. rate} \\
\cmidrule(lr){2-3}\cmidrule(lr){4-5}
 & $\rho$ & $p$ & $r_{pb}$ (95\% CI) & MW $p$ \\
\midrule
SMIT       & -0.17 & 0.011 & -0.07 [-0.21, 0.09] & 0.340 \\
Swin~UNETR & -0.17 & 0.011 & -0.15 [-0.30, 0.00] & 0.017 \\
\bottomrule
\end{tabular}
}
\end{table}

\printcredits

\bibliographystyle{cas-model2-names}

\bibliography{reference}

\end{document}